\documentclass[twocolumn,prX,showpacs,showkeys]{revtex4}

\usepackage{amsmath}
\usepackage{amsfonts}
\usepackage[english]{babel}

\newcommand{\mR}{\mathbb{R}}

\newcommand{\mS}{\mathbb{S}}

\newcommand{\cF}{\mathcal{F}}

\newcommand{\uX}{\underline{X}}
\newcommand{\uXb}{\widehat{\underline{X}}}

\newcommand{\uYb}{\widehat{\underline{Y} }}
\newcommand{\uY}{\underline{Y}}

\newcommand{\pj}{\partial_{x_j}}

\begin{document}
\title{Schr\"odinger equation with delta potential in superspace}

\author{Hendrik De Bie}
\affiliation{Clifford Research Group, Department of Mathematical Analysis, Faculty of Engineering, Ghent University, Galglaan 2, 9000 Gent, Belgium}
\email{Hendrik.DeBie@UGent.be}

\date{\today}

\begin{abstract}
A superspace version of the Schr\"odinger equation with a delta potential is studied using Fourier analysis. An explicit expression for the energy of the single bound state is found as a function of the super-dimension $M$ in case $M$ is smaller than or equal to 1. In the case when there is one commuting and $2n$ anti-commuting variables also the wave function is given explicitly.
\end{abstract}


\maketitle

\section{Introduction}

In recent work, we have been developing a new approach to the study of superspace, namely by means of harmonic and Clifford analysis (see e.g. \cite{DBS1,DBS4,DBS5,DBS3,DBS6}). The main feature of this approach, when working over the superspace $\mR^{m|2n}$, generated by $m$ commuting or bosonic variables $x_i$ and $2n$ anti-commuting or fermionic variables ${x \grave{}}_i$, is the introduction of a super Laplace operator $\Delta$ and a generalized norm squared $X^2$, defined by
\begin{eqnarray*}
\Delta &=&4 \sum_{j=1}^n \partial_{{x \grave{}}_{2j-1}} \partial_{{x \grave{}}_{2j}} -\sum_{j=1}^{m} \pj^2\\
X^2 &=& \uXb^2 +\uX^2 = \sum_{j=1}^n {x\grave{}}_{2j-1} {x\grave{}}_{2j}  -  \sum_{j=1}^m x_j^2.
\end{eqnarray*}

If we calculate the action of $\Delta$ on $X^2$, we find
\[
\Delta(X^2)= 2(m-2n)=2M.
\]

$M$ is the so-called super-dimension, a numerical parameter that characterizes several global features of the superspace $\mR^{m|2n}$ (see e.g. \cite{DBS5,DBS3}).

The aim of this paper is to show that this new framework can be used to study a Schr\"odinger equation on superspace, in the case where a delta potential is considered. This is physically interesting, because the Schr\"odinger equation in basic quantum mechanics is of course of the utmost importance. However, it is quite cumbersome to treat fermionic particles when only using commuting variables, as one has to anti-symmetrize the resulting wave functions in order to satisfy Pauli's exclusion principle. Due to the anti-commutativity of the variables ${x \grave{}}_i$, this problem is immediately tackled by the use of superspaces, which hence allow for an elegant and simultaneous description of bosonic and fermionic degrees of freedom.

On the other hand, most physical theories using superspaces and supermanifolds are dealing with highly theoretical problems, in subjects ranging from super-gravity to super-string-theory. It is thus very interesting and educational to see that also the theory of basic quantum mechanics is extendable to such superspaces, and that moreover the resulting spectra are given by replacing the ordinary Euclidean dimension $m$ by the super-dimension $M$ (see e.g. \cite{DBS3, ZHANG} and formula (\ref{energyeq})).

The classical Schr\"odinger equation on $\mR^{m|0}$ is given by (when working in units $m=\hbar=1$)
\[
-\frac{1}{2} \nabla^2 \psi(x_i) + V \psi(x_i) = E \psi(x_i).
\]
A canonical generalization of this equation to superspace is found by replacing $-\nabla^2$ by the super Laplace operator and $V$ by a potential in both the commuting and anti-commuting variables, yielding
\begin{equation}
\frac{1}{2} \Delta \psi(x_i;{x \grave{}}_j) + V(x_i ; {x \grave{}}_j) \psi(x_i ; {x \grave{}}_j) = E \psi(x_i; {x \grave{}}_j).
\label{superschrodinger}
\end{equation}
In this equation, $\psi(x_i;{x \grave{}}_j)$ is a super-function, i.e. a function of the general form
\begin{equation}
f = \sum_{\nu=(\nu_1,\ldots,\nu_{2n})} f_{\nu}(x_i) {x \grave{}}_{1}^{\nu_1} \ldots {x \grave{}}_{2n}^{\nu_{2n}}
\label{superfunction}
\end{equation}
where $\nu_i = 0$ or $1$ and $f_{\nu}(x_i)$ is a complex-valued function of the (real) co-ordinates $x_i$. 

This means that equation (\ref{superschrodinger}) is in general equivalent with a complicated set of partial differential equations in the $f_{\nu}(x_i)$, of which it is a priori not clear that it can be solved.
We can define a norm on superfunctions (\ref{superfunction}) e.g. as follows
\[
||f|| =  \sum_{\nu=(\nu_1,\ldots,\nu_{2n})} \int_{\mR^m} f_{\nu}(x_i) \overline{f_{\nu}(x_i)} dV(\uX).
\]

Equation (\ref{superschrodinger}) has already been studied in the case $V= -X^2/2$ (super harmonic oscillator, see e.g. \cite{MR830398,DELBOURGO} for the purely fermionic case and also \cite{DBS3}) and in the case $V = 1/X$ (super Kepler problem, see \cite{ZHANG}). In this paper we will study the case $V = \delta(X)$ with $\delta(X)$ the super Dirac distribution or delta potential. The equation will be solved using the general Fourier transform in superspace (see \cite{HDB1}).

The one dimensional version of this problem is well known and yields one bound state solution. Interesting enough, the problem is not solvable using Fourier analysis if one considers a higher dimensional generalization of the problem: the wave function is then singular at the origin, making it impossible to determine the corresponding energy.

However, our superspace approach allows us to study this problem not in higher but in smaller (i.e. more negative) dimensions $M$ by increasing the number of anti-commuting variables.
In the sequel, we will explicitly determine the energy of the bound state solution as a function of $M$. In the case $m=1$, $n \neq 0$ we will also give an explicit expression for the wave function.

\section{Schr\"odinger equation in superspace}

The one dimensional Schr\"odinger equation with a delta potential is given by
\[
-\frac{1}{2}\frac{d^2}{d x^2} \psi - a \delta(x)\psi = E \psi
\]
where $a$ is a positive constant, modeling the strength of the potential. It has one bound state solution, namely $\psi = e^{- a|x|}$ with corresponding energy $E=-a^2/2$.

A canonical generalization of this equation to superspace is given by (see equation (\ref{superschrodinger}))
\[
\frac{1}{2} \Delta \psi - a \delta(X) \psi = E \psi
\]
with the super Dirac distribution given by
\[
\delta(X) = \pi^{n} \delta(\uX) {x \grave{}}_{1} \ldots {x \grave{}}_{2n} = \frac{\pi^{n}}{n!} \delta(\uX) \uXb^{2n}
\]
where $\delta(\uX)$ is the classical Dirac distribution in $\mR^m$.

We will solve this equation using the super Fourier transform (see \cite{HDB1}), defined as
\[
\cF_{m | 2n}^{\pm} ( . )(Y) = (2 \pi)^{-M/2} \int_{\mR^{m|2n}} \exp{(\mp i \langle X , Y \rangle)}(. )
\]
with
\[
\langle X,Y \rangle =   - \sum_{i=1}^{m} x_i y_i +\frac{1}{2} \sum_{j=1}^{n}({x \grave{}}_{2j-1}{y \grave{}}_{2j} - {x \grave{}}_{2j} {y \grave{}}_{2j-1})
\]
and where $\int_{\mR^{m|2n}}$ is the Berezin integral (see \cite{MR732126}, \cite{DBS5}) defined by
\[
\int_{\mR^{m|2n}} f = \pi^{-n} \int_{\mR^m} dV(\uX) \partial_{{x \grave{}}_{2n}} \ldots \partial_{{x \grave{}}_{1}} f.
\]

We first put $E = -  \widetilde{E}$ as we wish to obtain negative energies. We then obtain, putting $F = \cF_{m | 2n}^{+}(\psi)$,
\[
-\frac{1}{2} Y^2 F - a (2 \pi)^{-M/2} \psi(0) = - \widetilde{E} F
\]
or
\[
F = \dfrac{a (2 \pi)^{-M/2} \psi(0)}{\widetilde{E} - Y^2/2}.
\]
In this equation $1/(\widetilde{E}-Y^2/2)$ equals
\begin{eqnarray*}
\dfrac{1}{\widetilde{E} - Y^2/2} &=& \dfrac{1}{r^2/2 + \widetilde{E} - \uYb^2/2}\\
&=&\dfrac{1}{r^2/2 + \widetilde{E}} \left( 1 - \frac{\uYb^2/2}{r^2/2 + \widetilde{E}}\right)^{-1}\\
&=&\dfrac{1}{r^2/2 + \widetilde{E}} \sum_{k=0}^{n} \frac{\uYb^{2k}/2^k}{(r^2/2 + \widetilde{E})^k}
\end{eqnarray*}
where $r^2 = -\uY^2 = \sum_i y_i^2$.
Applying the inverse Fourier transform $\cF_{m | 2n}^{-}$ to $F$ then yields
\begin{eqnarray*}
&&\psi(x_i;{x \grave{}}_j)\\&=& \cF_{m | 2n}^{-} (F)\\
&=& \frac{a \psi(0)}{(2 \pi)^{\frac{M}{2}}}  \sum_{k=0}^{n} \cF_{0 | 2n}^{-}\left(\uYb^{2k}/2^k \right) \cF_{m | 0}^{-}\left(\dfrac{1}{(r^2/2 + \widetilde{E})^{k+1}}\right)\\
&=& \frac{a \psi(0)}{(2 \pi)^{\frac{M}{2}}} \sum_{k=0}^{n}  \frac{2^{k-n} k!}{(n-k)!} \uXb^{2n-2k} \cF_{m | 0}^{-}\left(\dfrac{1}{(r^2/2 + \widetilde{E})^{k+1}}\right).
\end{eqnarray*}

The energy $\widetilde{E}$ is found by evaluating the wave function at $x_i={x \grave{}}_j=0$. This gives the following equation:
\[
1 = a  (2 \pi)^{-M/2}  n! \cF_{m | 0}^{-}\left(\dfrac{1}{(r^2/2 + \widetilde{E})^{n+1}}\right)_{\uX = 0}.
\]

This Fourier integral is evaluated as follows:
\begin{eqnarray*}
&&\cF_{m | 0}^{-}\left(\dfrac{1}{(r^2/2 + \widetilde{E})^{n+1}}\right)_{\uX = 0}\\ &=& (2 \pi)^{-m/2} \int_{\mR^m} e^{-i \langle \uX,\uY \rangle} \dfrac{dV(\uY)}{(r^2/2 + \widetilde{E})^{n+1}} |_{\uX = 0} \\
&=&(2 \pi)^{-m/2} \int_{\mR^m} \dfrac{dV(\uY)}{(r^2/2 + \widetilde{E})^{n+1}}\\
&=&(2 \pi)^{-m/2} \int_{r=0}^{+\infty} \dfrac{r^{m-1} dr}{(r^2/2 + \widetilde{E})^{n+1}} \int_{\mS^{m-1}} d \sigma\\
&=&\frac{1}{n!} \widetilde{E}^{M/2-1} \Gamma(1-M/2).
\end{eqnarray*}
Note that the integral over $r$ only converges if $M \leq 1$. This was to be expected, as the classical Schr\"odinger equation with delta potential is not solvable using Fourier methods in higher dimensions.

We hence obtain the following expression for the energy of the single bound state, as a function of the super-dimension $M$:
\begin{equation}
E = - \widetilde{E} = - \left( \dfrac{(2 \pi)^{M/2}}{ a \Gamma(1-M/2)} \right)^{\frac{1}{M/2-1}}.
\label{energyeq}
\end{equation}
Again, this expression is only well-defined if $M \leq 1$.

For $M$ tending to $- \infty$ we have the following asymptotic behaviour
\[
E \propto \frac{\pi}{e} M \approx 1.1557 M.
\]

In table \ref{energytable} we give the numerical value of the energy for several values of the super-dimension in the case $a=1$.

Let us now consider the case $m=1$, $n \neq 0$. In this case we can explicitly calculate $\cF_{1 | 0}^{-}((y^2/2 + \widetilde{E})^{-k-1})$ for all $k$ using contour integration and the residue theorem. This yields
\begin{eqnarray*}
&&\cF_{1 | 0}^{-}\left(\dfrac{1}{(y^2/2 + \widetilde{E})^{k+1}}\right)\\&=& \sqrt{2 \pi} \frac{2^{k+1}}{k!} e^{-\sqrt{2\widetilde{E}} |x|} \sum_{j=0}^{k} \frac{(k+j)!}{j! (k-j)!}\frac{|x|^{k-j}}{(2 \sqrt{2\widetilde{E}})^{k+j+1}}.
\end{eqnarray*}
\begin{widetext}
The complete wave function is then given by
\[
\psi(x) = a  \pi^{n} \psi(0)  e^{-\sqrt{2\widetilde{E}} |x|} \sum_{k=0}^{n} \sum_{j=0}^{k}  \frac{2^{2k+1}}{(n-k)!}   \frac{(k+j)!}{j! (k-j)!}\frac{|x|^{k-j}}{(2 \sqrt{2\widetilde{E}})^{k+j+1}} \uXb^{2n-2k}.
\]
\end{widetext}
In principle, $\psi(0)$ can be determined by normalizing the wave function such that $||\psi(x)||=1$. This leads however to a very complicated formula. In the case $m=1$, $n=0$ everything reduces to the classical result.

In general dimension the complete wave function is found by calculating $\cF_{m | 0}^{-}\left(r^2/2 + \widetilde{E}\right)^{-k-1}$. Explicit expressions can be found e.g. in \cite{GELFAND}. Note that the inverse Fourier transform is singular at the origin if $m \geq 2k+2$. However, this poses no problems as long as the body of the wave function (i.e. the function obtained by putting all anti-commuting variables equal to zero) is finite at the origin. This is the case if $M \leq 1$.

\begin{table}
	\caption{The energy as a function of the super-dimension $M$ ($a = 1$).}
		\begin{tabular}{c|c}
			M& Energy\\
			\hline\\
			1& -0.5\\
			0& -1\\
			-1&-1.7025\\
			-2&-2.5066\\
			-3&-3.3757\\
			-1000&-1148.8
			\end{tabular}
	\label{energytable}
\end{table}

\section{The purely fermionic case}

In the purely fermionic case, only anti-commuting variables are considered (i.e. $m=0$, $n \neq 0$). 
First note that solving the Schr\"odinger equation is then a purely algebraic eigenvalue problem.

The method of the previous section can be used to determine the energy level
\[
E= - \left(a n! (2\pi)^n \right)^{\frac{1}{n+1}}
\]
and the corresponding wave function. However, if $n$ is odd, we also have the level $E = \left(a n! (2\pi)^n \right)^{\frac{1}{n+1}}$, corresponding to the negative root from equation (\ref{energyeq}). Moreover, there is always an energy level $E = 0$ with degeneracy $\binom{2n+1}{n}-1$.
This can easily be proven using the general method developed in \cite{HDB3} to solve the fermionic Schr\"odinger equation.

\section{Conclusions}
In this paper we have defined a Schr\"odinger equation on superspace and solved it using Fourier analysis in the case where the potential is the delta potential and the super-dimension is smaller than or equal to 1. We have obtained an explicit expression of the energy of the single bound state solution in terms of the super-dimension. In the case where there is only one commuting variable we have also completely determined the wave function.

\begin{acknowledgments}
The author is a Ph.D. Fellow of the Research Foundation - Flanders (FWO).
\end{acknowledgments}

\end{document}